\documentclass[
aip,
jasa,
amssymb,
reprint,
showpacs]{revtex4-1} 

\usepackage[pdftex]{graphicx}
\usepackage{amsmath}
\usepackage{chemformula}

\usepackage{sidecap}
\sidecaptionvpos{figure}{t}
\usepackage[colorlinks=true,linkcolor=blue,citecolor=blue,filecolor=blue,urlcolor=blue]{hyperref}

\usepackage{tabularx}
\newcolumntype{A}{>{\centering\arraybackslash}X}
\usepackage{color}

\begin{document}
\title[Monitoring the setting of calcium sulphate bone-graft substitute using ultrasonic backscattering]{Monitoring the setting of calcium sulphate bone-graft substitute using ultrasonic backscattering}      

\author{Josep Rodr\'\i{}guez-Sendra}
\email{jorosen@upv.es}
\affiliation{I3M, Instituto de Instrumetaci\'on para Imagen Molecular, CSIC - Universitat Polit\`ecnica de Val\`ecia, Valencia, Spain}%
\author{No\'e Jim\'enez}
\email{nojigon@upv.es}
\affiliation{I3M, Instituto de Instrumetaci\'on para Imagen Molecular, CSIC - Universitat Polit\`ecnica de Val\`ecia, Valencia, Spain}%
\author{Rub\'en Pic\'o}
\affiliation{Universitat Polit\`ecnica de Val\`ecia, Valencia, Spain}%
\author{Joan Faus}
\affiliation{Insitituto Valenciano de Investigaciones Ondontol\'olicas, Gandia, Spain}%
\author{Francisco Camarena}
\affiliation{I3M, Instituto de Instrumetaci\'on para Imagen Molecular, CSIC - Universitat Polit\`ecnica de Val\`ecia, Valencia, Spain}%
\thanks{corresponding author}

\date{\today}

\begin{abstract}
	We report a method to monitor the setting process of bone-graft substitutes
(calcium sulphate) using ultrasonic backscattering techniques. Analyzing the backscattered fields using a pulse-echo technique, we show that it is possible to dynamically describe the acoustic properties of the material which are linked to its setting state. Several experiments were performed to control the setting process of calcium sulphate using a 3.5-MHz transducer. The variation of the apparent integrated backscatter (AIB) with time during the setting process is analyzed and compared with measurements of the speed of sound (SOS) and temperature of the sample. The correlation of SOS and AIB allows to clearly identify two different states of the samples, liquid and solid, in addition to the transition period. Results show that using backscattering analysis the setting state of the material can be estimated with a threshold of 15 dB. This ultrasonic technique is indeed a first step to develop real-time monitoring systems for time-varying complex media as those present in bone regeneration for dental implantology applications.
\end{abstract}

\maketitle

\section{Introduction}\textsl{}
Dental implants require sufficient bone volume to achieve stability and durability. The loss of tooth affects the quality and thickness of the bone \cite{Bodic2005}. In cases where bone volume is insufficient, guided bone regeneration (GBR) techniques are applied\cite{Bodic2005}. In particular, GBR by maxillary sinus floor augmentation procedures are used since 50 years, and were first proposed by Tatum in 1974 \cite{Tatum1986}. In this operation, a window in the lateral area of the gingival is performed, the Schneider membrane is lifted, and a bone graft inserted. The window is coated with a pericardium membrane. Full bone regeneration process has a typical healing duration of approximately 6 months. The procedure sometimes fail\cite{Annibali2008} and different complications may appear postoperatively, e.g. soft tissue growth or failed osseointegration \cite{JAcero2011}. Odontologists make use of X-ray radiological techniques in order to evaluate the bone regeneration process \cite{memon2010dental}.
However, monitoring the full regeneration process using these techniques imply potential risks, such as excessive exposure to ionizing radiation \cite{memon2010dental,Ludlow2008}. In this sense, non-invasive ultrasonic techniques are desirable to monitor bone-regeneration during the healing time since they are non-ionizing.

Ultrasound has been used fruitfully to characterize the physical properties of bone. Technology for bone density estimation is nowadays mature \cite{Abendschein1970,Rossman1989}. Many commercial devices exist in the market used for the diagnosis of osteoporosis, e.g., for calcaneus or phalanx \cite{Langton1984,Langton2008}. Acoustic and elastic parameters are correlated with the physical properties of bone for diagnostic purposes. Among them, the speed of sound (SOS) is the most spread \cite[Chapter 3]{Laugier2011}. It consist in estimating the effective travelling velocity of a pulsed excitation, associated to the group speed of longitudinal waves in a given frequency band. Bone tissue is a heterogeneous biphasic medium composed of a multiple-scale porous viscoelastic matrix, soft-solids and viscous fluids \cite{Laugier2011}. Acoustic propagation through these complex media is therefore strongly dispersive and SOS, given as a single value, should be interpreted as an effective parameter rather than an intrinsic physical parameter locally related to the mechanical properties of the bone tissue. A simple way to measure the SOS of bone is to estimate the time-of-flight difference between two received echoes using a sample of known thickness. 

In addition, ultrasonic waves travelling in bones are strongly scattered in a diffuse way due to their complex heterogeneous micro-structure. In this way, the interference of multiple scattered waves from internal micro-structures can be received at source position. During last two decades, different techniques based on the backscatter energy evaluation have been proposed for bone characterization \cite{Wear1998,Chaffai2002}. Ultrasonic backscatter energy can be linked to the physical properties of bone tissue such as bone volume fraction or bone mineral density \cite{Laugier2011,padilla2008,karjalainen2009}, enabling quantitative ultrasound techniques for bone characterization. 

Many ultrasonic backscattering techniques are based on measurements of the apparent backscatter transfer function (ABTF)\cite{Laugier2011}. It represents the backscattered power from the sample corrected for the frequency response of the measurement system. Two parameters without explicit dependence of frequency are obtained from ABTF. The apparent integrated backscatter (AIB) is one of the most spread. It consist of a measure of the frequency-averaged (integrated) backscattered power contained in some portion of a backscattered ultrasonic signal \cite{Hoffmeister2006}. The frequency slope of apparent backscatter (FSAB) is determined by fitting a line in the frequency domain to the spatially averaged backscattered ultrasonic signals over the analysis bandwidth \cite{Hoffmeister2011}. Both AIB and FSAB are determined in a logarithmic scale and are referenced to the measured reflected energy of a reference flat panel, e.g., a steel sample. In order to remove this reference, Hoffmeister proposed in \cite{Hoffmeister2012} a set parameters that compare the backscatter energy from two different parts of the sample. The mean of the backscatter difference spectrum (MBD) is obtained by frequency averaging the difference between the backscatter spectrum over the analysis bandwidth from two different windows of the ultrasonic signal.

During bone regeneration process, its mechanical properties are modified from the initial bone graft to the consolidated bone. A change of state is produced as a consequence of osseointegration. Thus, a solution for detecting possible anomalies during the healing of bone in GBR applications consist of monitoring the ultrasonic properties of tissue during time. In this sense, ultrasound techniques have been applied for characterization of slowly time varying materials. One relevant example are cements, that have been characterized by ultrasonic methods for more than 80 years, see e.g, \cite{naik2003ultrasonic,philippidis2005experimental}. In particular, pulse-echo ultrasonic techniques can be used to characterize the time dependency of SOS and the acoustic impedance during their setting process\cite{keating1989comparison,sayers1993ultrasonic,boumiz1996mechanical}. Cement starts to set when mixed with water and a series of hydration chemical reactions are produced. The constituents slowly hydrate and the mineral solidifies. In particular, the typical setting duration time for bone-graft materials based in calcium sulphate ranges from 30 minutes to 60 minutes\cite{Carlson2003,vlad2012ultrasound}. Same as cement setting, the GBR is a process in which the mechanical (and acoustical) properties of the material vary with time and a change of state is produced. The advantage of using a bone graft substitute as a phantom for monitoring GBR is that the duration of the process is reduced from 6 months-time of healing, to less than an hour in a controlled environment.

In this work, we propose a proof of concept to monitor the GBR by sinus augmentation using ultrasonic backscattering methods. Thus, the time-varying acoustical properties of the material under inspection are obtained using an echo-impulse technique. In particular, we use a phantom of calcium sulphate\cite{finkemeier2002bone,wang2017bone}, which has already been used for GBR in the past \cite{DeLeonardis2000,Thomas2009}. The change of state from a viscous fluid to a porous solid during the setting mimics the GBR in a much shorter duration that can be studied under laboratory conditions. We report that ultrasonic backscatter energy can be linked to the physical properties of tissue. In GBR applications the transmission mode using a pair of ultrasonic transducers is not allowed as a receiver transducer can not be placed behind the area under inspection. Thus, we propose the use of a single sensor improving the accessibility of medical instrumentation for tissue characterization.

\section{Methods}
\begin{figure}[t]
	\centering
	\includegraphics[width=1\linewidth]{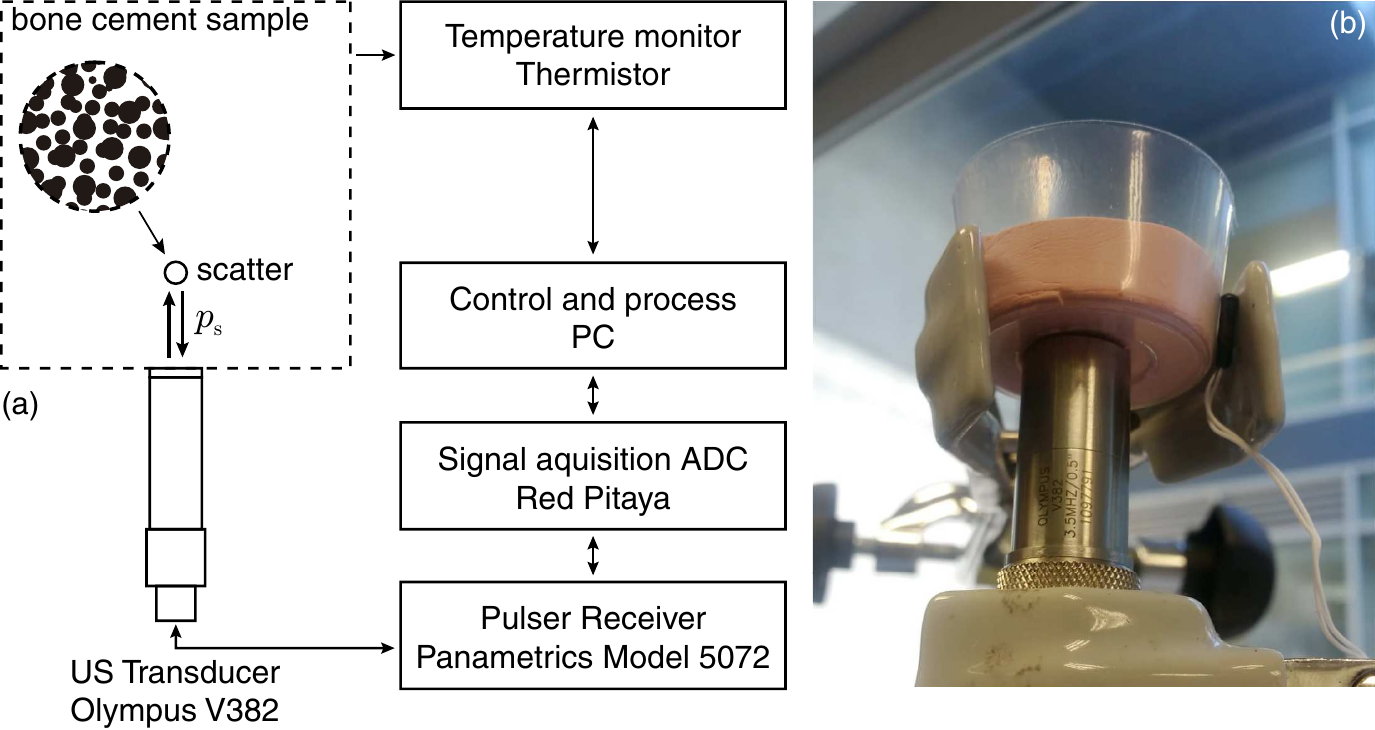}
	\caption{(a) Block diagram of the experimental setup. (b) Photograph of one experiment.}
	\label{Disp exp}
\end{figure}

\begin{figure*}[t]
	\centering
	\includegraphics[width=\textwidth]{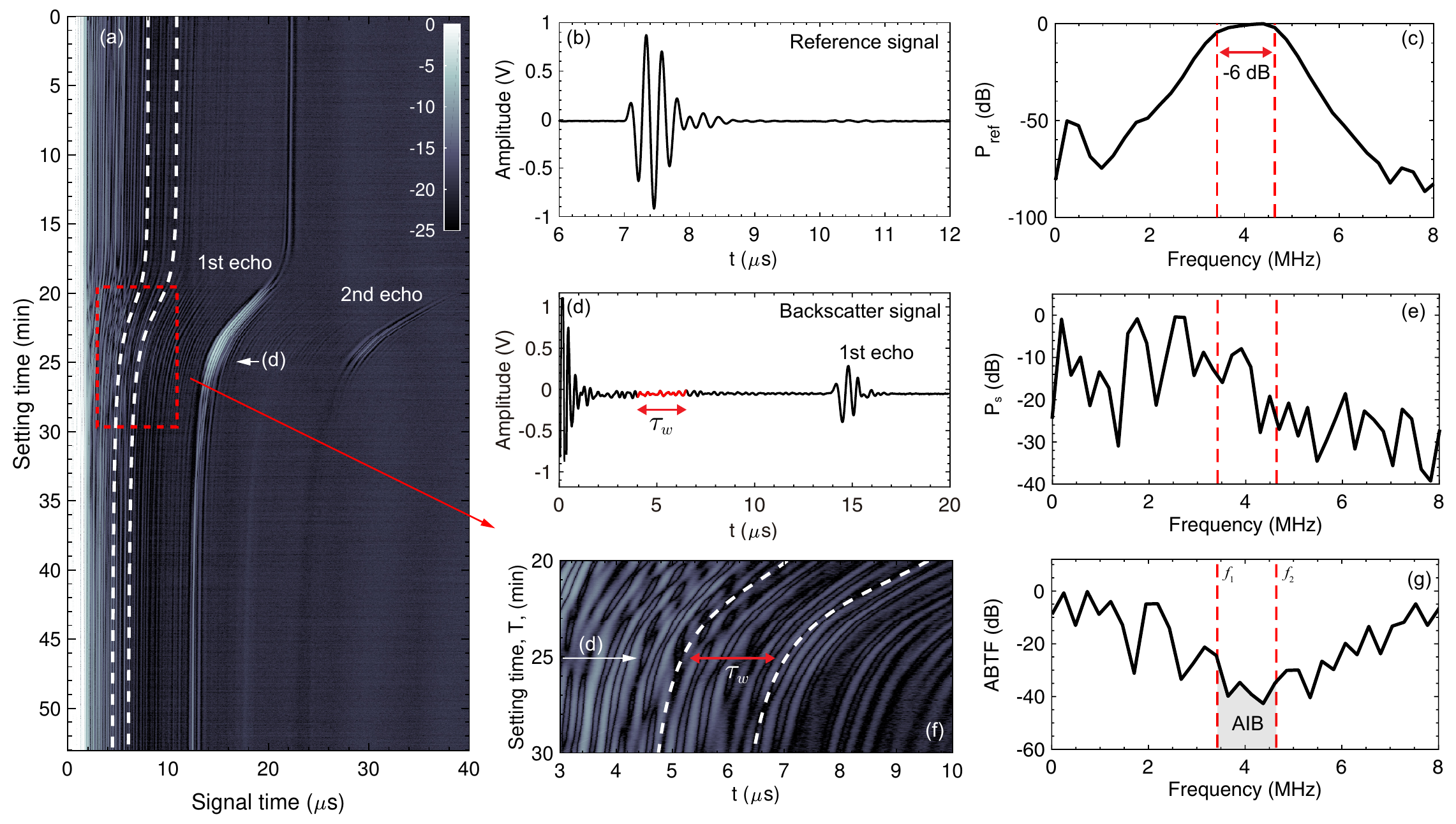}
	\caption{(a) Representation of all the experimental signals (RAW data) as a function of the setting time. Colormap in normalized dB scale. The white dashed lines mark the temporal window used to calculate the apparent integrated backscatter energy (AIB). (b) Reference signal (b) and (c) corresponding spectrum. (d) Example of a backscatter signal measured at $t=25$ min and (e) corresponding spectrum. (f) Zoom of the experimental data over the transition period. (g) Apparent backscatter transfer function (ABTF) and representation of the apparent integrated backscatter energy (AIB) (shaded area). }
	\label{fig:referencesignal}
\end{figure*}
\subsection{Phantom preparation}
Calcium sulphate type IV (Ventura Pinkmod) is the synthetic bone substitute used to perform the experiments. In particular, in its powder state the material is calcium sulfate hemihydrate (\ch{CaSO4 . 1/2 H2O}) in its $\alpha$ form. When the hemihydrate is mixed with water, calcium sulphate dihydrate is formed in a moderate exothermic reaction as\cite{Thomas2009}: 
\begin{center}
\ch{CaSO4 . 1/2 H2O + 1_1/2 H2O -> CaSO4 . 2 H2O + Heat}.
\end{center}

In this way, during the setting process the material shows a change of phase from a viscous liquid to porous solid. The total setting time is around 20 minutes. To study repeatability, four experiments were performed with samples composed of in 23 ml of water and 50 g of calcium sulphate. The powder was dissolved in water at 20 $\pm 0.2^\circ$ C and  mixed up to get homogeneous samples. The material was deposited in a cylindrical plastic container with a diameter of 36 mm to provide samples of 18 mm of thickness. 

\subsection{Ultrasonic measurement}
An Olympus V382 transducer with a nominal central frequency of 3.5 MHz and a bandwidth of 2.34 MHz (-6 dB) was used as emitter and receiver using a pulse-echo technique. Figure \ref{Disp exp} shows the block diagram and a photograph of the experimental setup. The transducer was placed in contact to the bottom of the recipient with the sample. Vaseline was applied between the transducer and the container to enhance acoustic coupling, and between the container and the sample to avoid decoupling during the solidification process. An ultrasonic pulser-receiver (Panametrics Model 5072) was used for emitting and receiving the acoustic pulses. The received signal was digitized with a data acquisition platform (Red Pitaya) at a sampling frequency of 125 MHz. For each experiment, a total of 1600 acquisitions were performed, each one every 2 seconds. The total duration of each experiment took 53 minutes, enough to cover the full setting process of the samples. Temperature was registered with thermistor probe (Tinytag Temperature Logger TK 4023) located externally in contact with the container. Measurements were post-processed to provide the acoustic parameters during the setting process.

\subsection{Speed of sound }
The speed of sound (SOS)  was calculated as the distance ratio between two times the thickness of the sample and the time of flight between the first two echoes from the end of the sample as
\begin{equation}\label{eq_c}
c =\frac{2L}{\Delta t},
\end{equation}
where $L$ is the thickness of the sample and $\Delta t$ is the time of flight, estimated as the times corresponding to the peak value of first two echoes. SOS was estimated during setting time, and a purely phenomenological model was fitted. The proposed model describes a smooth and asymmetric transition between two phases, fluid and solid, and is given by:
\begin{equation}\label{eq_arc}
c(t) = c_1 + (c_2-c_1)\tan^{-1}\left[\frac{\gamma }{\pi}(t-t_m) +\frac{1}{2}\right]^\beta,
\end{equation}
where $c_1$ is the speed of sound in liquid state, $c_2$ is the speed of sound in solid phase, $\gamma$ models the speed of the transition from liquid to solid phase, $t$ is the setting time, $t_m$ is the half-time of the setting process and $\beta$ is related to the symmetry of the function. Using this model we can compare the setting process of several experiments. 

\subsection{Backscatter parameters}
Two quantitative parameters based on the backscatter energy were considered in this work: the apparent backscatter transfer function (ABTF) and the apparent integrated backscatter energy (AIB). Both have been proven to be reliable and robust to characterize bone tissue\cite{Hoffmeister2015}. On the one hand, the ABTF is given by the transfer function between the backscatter signal and a reference signal as
\begin{equation}\label{eq_ABTF}
ABTF = 10\log_{10}\left(\frac{P_s(f)}{P_\mathrm{ref}(f)}\right),
\end{equation}
where $P_s(f)$ is the frequency-dependent power of a backscatter signal $p_s(t)$ using a time window of duration $\tau_w$, and $P_\mathrm{ref}(f)$ is the frequency-dependent power of the first echo from a reference reflector, e.g., steel plate \cite{Hoffmeister2015}. On the other hand, the AIB is obtained as the frequency-averaged ABTF as
\begin{equation}\label{eq_AIB}
AIB = \frac{1}{\Delta f}\int\limits_{f_1}^{f_2}ABTF(f)df,
\end{equation}
where $\Delta f = f_2-f_1$, and $f_1$ and $f_2$ are given by the -6 dB bandwidth of the transducer. The reference pulse and its spectrum are shown in Figs.~\ref{fig:referencesignal}~(b,c), respectively. The corresponding values for the effective limiting frequencies are $f_1=3.4$ MHz and $f_2=4.6$ MHz.

\begin{figure}[t]
	\centering
	\includegraphics[width=\columnwidth]{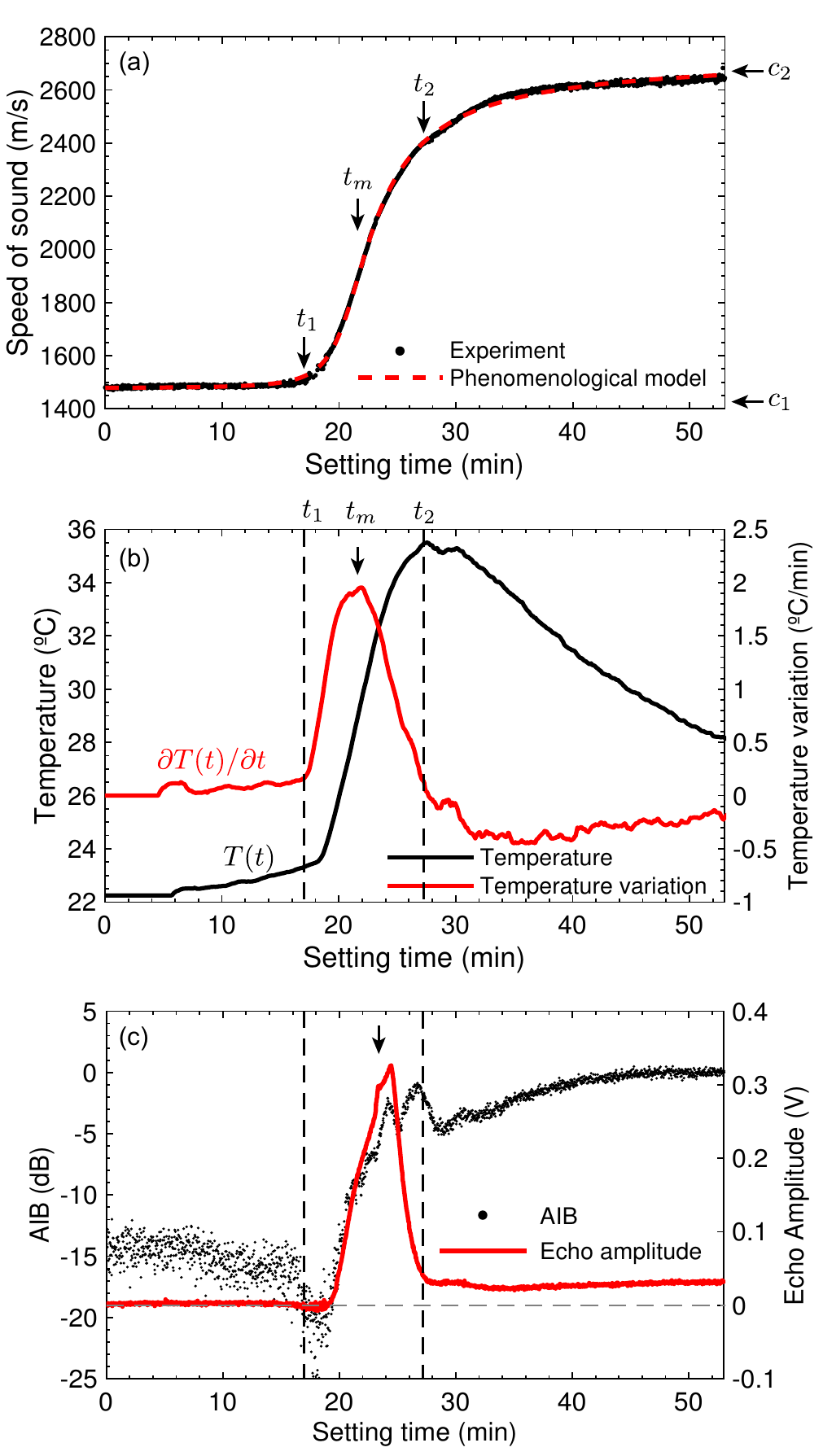}
	\caption{(a) Evolution of the speed of sound (SOS), experiment (dots) and fitted model (red dashed). (b) Evolution of the temperature (black) and temperature variation (red). (c) Apparent integrated backscatter energy (AIB) (dots) and peak amplitude of the first echo (red), as a function of the setting time.}
	\label{fig:aib-amplit}
\end{figure}

\begin{table*}
	\begin{center}
		\caption{Fitted parameters for the phenomenological model given by Eq.~(\ref{eq_arc}).}
		\begin{tabular*}{\textwidth}{l @{\extracolsep{\fill}} ccccc}
			\hline 
			& Transition asymmetry & Transition speed & Transition time & Initial state SOS & Final state SOS \\ 
			& $\beta$ & $\gamma$ (min$^{-1}$) & $t_m$ (min) & $c_{1}$ (m/s)& $c_{2}$ (m/s)\\ 
			\hline 
			Sample 1 (E1) &2,1  &0,346  &20,7  &1477,2  &2731,5  \\ 
			Sample 2 (E2)  &1,9  &0,341  &20,8  &1399.7  &2490.9  \\ 
			Sample 3 (E3)  &1,4  &0,327  &19,9  &1553,9  &2919,5  \\ 
			Sample 4 (E4)  &1,7  &0,367  &20,1  &1627,2  &3049,8  \\ 
			Mean  &$1,8\pm0.3$  &$0,345\pm0,016$  &$20,4\pm0.4 $ &1514.5$\pm$98.0  &$2858.1\pm 155.5$ \\ 
			\hline 
		\end{tabular*} 
		\label{tab:par}
	\end{center}
\end{table*}

\section{Results}
\subsection{Description of backscatter signals}
The experimental signals of one experiment are shown in Fig.~\ref{fig:referencesignal}. First, in Fig.~\ref{fig:referencesignal}~(a) the A-lines are presented as a function of the setting time. Here, the amplitude of the ultrasonic signal registered by the transducer is represented in logarithmic scale normalized to the maximum value for visualization. The vertical axis represents the time of the experiment during the setting process (in minutes) and the horizontal axis represents the time of the ultrasonic signal (in $\mu$s). As both scales are very different ($\sim10^6$), they are decoupled and each ultrasonic capture is assumed to be instantaneous.

\begin{figure}[t]
	\centering
	\includegraphics[width=0.86\columnwidth]{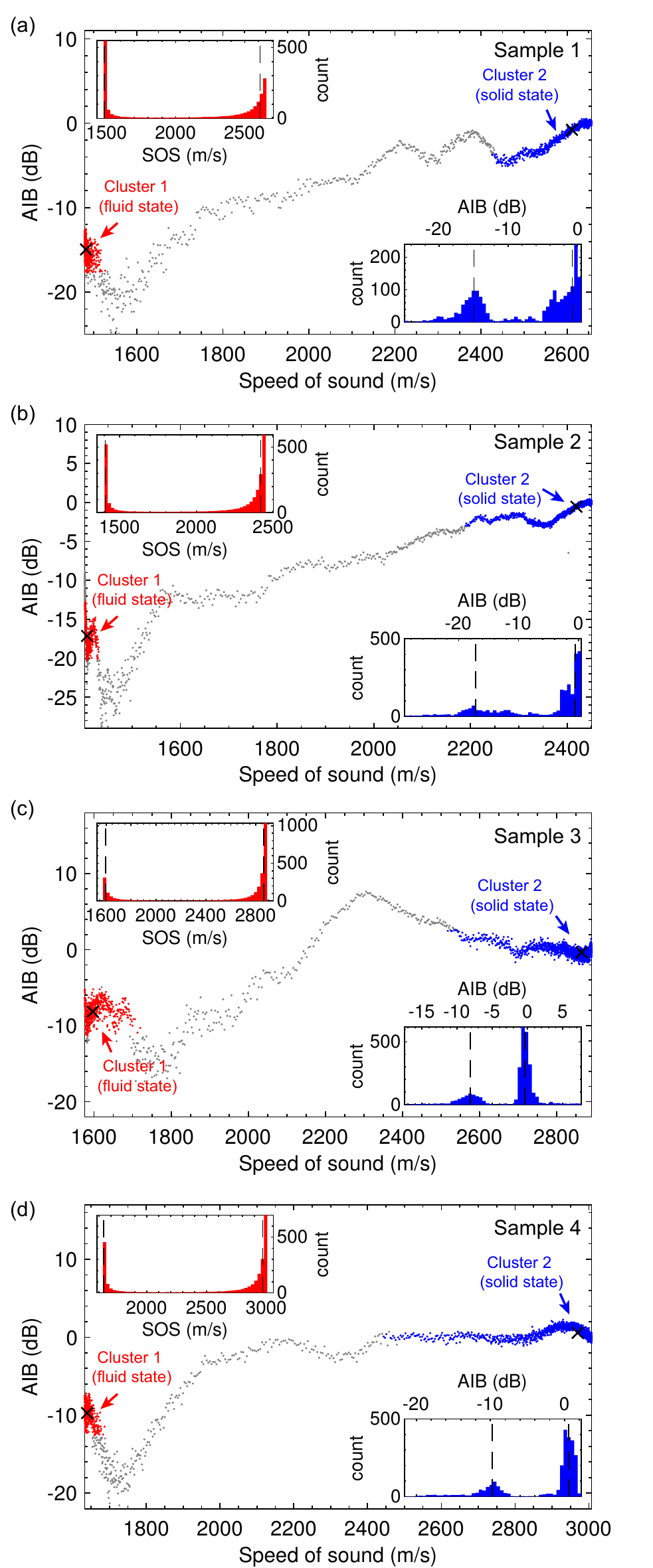}
	\caption{(a-d) Correlation of AIB and SOS and clustering for the four samples. The insets show the corresponding histograms of the SOS (red) and AIB (blue). Each detected cluster is shown in color (red and blue), corresponding to a different state. The mean value of each cluster is marked by black crosses.}
	\label{fig:velaib-corr}
\end{figure}

From $t = 0$ $\mu$s to $t = 2$ $\mu$s, all ultrasonic signals have a maximum value of 0 dB corresponding to the emission pulse. Then, vertical grey traces appear corresponding to the received backscattered waves. First, when the experiment is started, the sample presents a viscous fluid state. At the beginning of the experiment time, at the top of Fig.~\ref{fig:referencesignal}~(a), the echo from the end of the sample is visible at $t = 23$ $\mu$s. Up to about 15 min, the backscatter signals are stable and only small changes are visible, as shown by the straight vertical traces in the map. However, from about 18 min to 27 min a sudden and drastic change in the signal is observed: the line corresponding to the echo from the back of the sample is shift towards the beginning of the signal. This temporal compression of the signal is related to the increase of SOS in the material: the echo from the back is arriving sooner, as we will see after. Then, at 27 min the signals become stable again and, therefore, SOS should present a stable value at the end of the experiment. At 50 min the sample is physically a porous solid material. 

Dashed lines in Fig.~ \ref{fig:referencesignal}~(a) show the limits for the temporal window used for the post-processing of the backscattering. For their estimation, both the initial and final temporal values were corrected using the SOS in order to capture a constant spatial window. Figure~\ref{fig:referencesignal}~(f) shows in detail the tracking of lines during the transition from liquid state to solid state in the setting process. Here, the traces of the captured backscattered waves match the analysis window marked by the dashed white lines. 

Backscatter parameters have been referenced to a signal obtained from a flat steel plate, shown in  Fig.~\ref{fig:referencesignal}~(b). An example of A-scan signal registered by the transducer using the bone graft substitute sample (at $t=25$ min) is shown in Fig.~\ref{fig:referencesignal}~(d). Both the initial electrical contribution and the first echo caused at the end of the sample can be observed at $t = 0$ $\mu$s and $t = 15$ $\mu$s, respectively. In comparison, backscattered energy is very low and spreads all along the signal. We consider a window of duration of $\tau_w$ = 2.8 $\mu$s, i.e., 10 times the period corresponding of the central frequency of the transducer at the beginning of the experiment, according to \cite{Hoffmeister2006}. The window was set initially centred at 8 $\mu$s. Note that this value is shifted at 25 mins (Fig.~\ref{fig:referencesignal}~(d)) to 5 $\mu$s due to the change of SOS with the setting time (see white dashed lines in Fig.~\ref{fig:referencesignal}~(a)). The spectra of the reference signal and the example backscattered signal are shown in Figs.~\ref{fig:referencesignal}~(c,e), respectively. Then, for each temporal signal ABTF is calculated using Eq.~(\ref{eq_ABTF}), as shown in Fig.~\ref{fig:referencesignal}~(f). Finally, AIB is estimated using Eq.~(\ref{eq_AIB}) by averaging ABTF over the interest bandwidth, shown in Fig.~\ref{fig:referencesignal}~(f) as the shaded region.

\subsection{Evolution of speed of sound}
The evolution of the SOS with the setting time is shown in Fig.~\ref{fig:aib-amplit}~(a). In addition, the evolution of the temperature and its variation, i.e., $\partial T/\partial t$, are shown in Fig.~\ref{fig:aib-amplit}~(b). The initial value measured for the SOS corresponds to $c_1= 1477.2 \pm 50$ m/s close to the value of liquid water at same temperature (1495 m/s). Up to $t_1=17.32$ min, temperature is approximately stable and its rate of variation is small, about $0.1^\circ$C/min. However, after this time ($t_1$), the chemical reaction between the calcium sulfate and water suddenly increases. When temperature starts to increase small variations of SOS were already detected.

The maximum temperature variation is achieved at 21.6 min, corresponding to $1.9^\circ$C/min. This point corresponds to the maximum exothermic chemical activation\cite{liu2003exothermal}. Note this time almost corresponds to the time $t_m=20.7$ min given by the SOS model fit. After that moment, there is more consolidated material than calcium sulphate hemihydrate and water available to maintain the chemical reaction at this rate and, therefore, the exothermic activity is smoothly reduced. Then, at $t_2=27.6$ min the temperature reaches its maximum value, $35.5^\circ$C. The sample starts to cool smoothly due to heat diffusion processes and the lack of chemical activity. Once the change of state is completed, the sample is transformed from a viscous liquid to porous solid material. When the material is solidified, the SOS reaches a stable value close to $c_2=2731.5$ m/s. The behaviour of this synthetic bone substitute agrees the one reported for other cements measured using a emission/reception mode\cite{keating1989comparison,sayers1993ultrasonic,boumiz1996mechanical}. 

As expected, it can be observed that the change of the state of the material corresponds to an abrupt increase of the SOS. This particular smooth transition was measured in all samples, and can be described by three different phases: the liquid state, the transition and the solid state. The experimental SOS data was fitted to the phenomenological model given by Eq.~(\ref{eq_arc}). The parameters of the fitted model describe quantitatively the setting process. These parameters are listed in Table~\ref{tab:par} for the 4 samples, showing similar values between experiments. 

First, the parameter $c_1$ is linked to the physical SOS at the initial viscous-liquid state of the sample. The mean value of $c_1$ over the four experiments is 1514.5 $\pm$ 98.0 m/s. As the material at the initial process is mainly water with undissolved powder, the SOS measured for the sample is close to the SOS of the water. The final value of the SOS in the experiment is related to $c_2$, i.e, the SOS of the material in solid state. The mean value of $c_2$ is 2858.1 $\pm$ 155.5 m/s. 

The transition time for the setting process is given by the parameter $t_m$. This parameter takes a mean value of $t_m=20.4$ min, very close to the peak of maximum exothermic chemical activity (21.6 min for sample 1). The factor $\gamma$ is related to the duration of the transition from liquid to solid state, its mean value is 1.8 $\pm$ 0.3 min$^{-1}$. Finally, the factor $\beta$ is related to the symmetry of the function and takes a mean value of 0.345 $\pm$ 0.016. The dispersion of the parameters is low, and all experiments present similar transition curves. However, some discrepancies are observed in SOS values, that can be associated to uncertainness on estimating the thickness of the sample $L$.

\subsection{Evolution of apparent integrated backscatter energy}

In addition to SOS, energetic parameters also can provide information about the change of phase of calcium sulphate. In particular, we show in Fig.~\ref{fig:aib-amplit}~(c) the AIB as a function of the setting time. Moreover, the amplitude of the 1st echo is also shown by the red curve in Fig.~\ref{fig:aib-amplit}~(c). When the sample is in a viscous-liquid state, from $t=0$ to about $t_1$, the mean value of AIB is low and constant, taking a value of about $-15$~dB. However, the AIB estimations present high dispersion due to the low amplitude of the backscatter signals. Note that, in addition, the amplitude of the 1st echo is also very low. The weak scattering and the weak echo reflection are caused by the strong absorption and high homogeneity of the sample in its initial viscous-liquid phase. When the exothermic chemical reaction starts, both AIB and echo signal initially decrease. AIB takes a value down to -20 dB at 20 min. 

However, as the reaction continues (after minute 17) both the scattering energy and the peak amplitude of the echo increase: the intrinsic absorption of the consolidated synthetic bone substitute is lower compared to its viscous-fluid phase. Thus, AIB progressively increases and its dispersion is reduced. During this time the transition of state in the setting process is produced; the mechanical and acoustic features of the sample are drastically modified and the microstructure of the porous solid material starts to consolidate, leading to a increase of the amplitude of the backscatter signal. The consolidated material present less absorption, and at the middle of the transition the echo from the back arrives with remarkable amplitude. However, as the porous solid consolidates its inner microstructure the backscatter energy increases. As a consequence, the amplitude of the 1st echo of the back is again reduced as waves reach the end of the sample with lower amplitude. After some small oscillations during the transition the AIB ends with a value of about -5 dB at 25 min. 

Finally after 30 min, the sample is transformed to solid state and the AIB is smoothly increased from -5 dB to a plateau reaching a maximum value 0 dB at 50 min. Therefore, while the dispersion of the AIB is low and it reaches a stable value, the amount of the energy of the ultrasonic pulse that reaches the end of the sample is reduced due to scattering processes. This is caused by the final properties of the bone graft substitute that present high micro-heterogeneity and porosity. After this time, the setting process is completed and the material is mechanically and acoustically stable.

\subsection{Correlation between of SOS and AIB}

While at the initial and the end of the setting time AIB takes different values, AIB presents some oscillations during the setting period. These can lead to uncertainness if used to monitor the setting of the material. A deeper analysis can be performed by showing the correlation between AIB and SOS. Figures~\ref{fig:aib-amplit}~(a-d) show this correlation for the four experiments. For each experiment, the insets show the corresponding histograms for SOS and AIB data, where two peaks corresponding to the initial and final states can be observed. While there exist a positive correlation between AIB and SOS, the data show some oscillations. 

A concentration of experimental points can be observed in two zones, marked by arrows. On the one hand, when the material is in its viscous liquid phase is when the values of SOS are minimum. On the other hand, when the material is a porous solid both SOS and AIB present maximum values. As experimental data concentrates in these two limits, we can apply clustering techniques to identify them. A density-based spatial clustering of applications with noise (DBSCAN) technique\cite{ester1996density} was employed here to discern between the data corresponding to the two concentration points and the transition. This give us the corresponding clusters marked in red and blue in Figs.~\ref{fig:aib-amplit}~(a-d), helping to monitor the setting process: when a clusters is clearly populated, the sample has reached a stable phase. Note that using only the histogram for the AIB some of the signals corresponding to the transition point can present the same AIB than in the final state due to the oscillations of this parameter, see e.g. Fig.~\ref{fig:velaib-corr}~(c).

\section{Conclusions}

The change of state of a bone graft substitute (calcium sulphate) has been monitored and characterized by ultrasonic backscattering analysis. Temperature, speed of sound (SOS) and apparent integrated backscatter energy (AIB) were measured during the setting process of the sample using a pulse-echo technique. The evolution of SOS shows a smooth transition between two limiting values, corresponding to the viscous-fluid and porous solid state of the calcium sulphate samples. An excellent agreement is found between the evolution of the SOS and a simple phenomenological model to describe the setting process. The experiments also show good repeatability. 

In addition, the evolution of AIB is shown to be correlated with SOS. The apparent integrated backscatter energy is proven to be an appropriate parameter to describe the transition from a viscous-liquid state, with values around -15 dB, to a complex porous solid, with values around 0 dB. This relation confirms that the backscattering analysis with ultrasound is a powerful tool to monitor complex materials with time-varying mechanical properties using a single transducer and pulse-echo techniques.

This study represents a contribution to the use of ultrasound for the monitoring of GBR, which is necessary for the successful placement of dental implants. A synthetic bone substitute has been used as a bone phantom because its mechanical properties change during the setting process from a viscous liquid to a rigid porous material, as, under reasonable simplifications, occurs in GBR applications. However, many practical questions remains open and should be explored. These include the effect of the varying coupling of the transducer and the tissue between acquisitions, the mechanical/acoustical behaviour of real tissue during bone regeneration, the scattering properties of these unconsolidated and consolidated tissues, patient variability, or the impact of the intrinsic properties of real bones during regeneration (porosity, tortuosity, viscosity, ...) in the estimated parameters. These issues should be explored in future studies. 

Finally, it is worth to mention that one advantage of backscattering analysis with respect to speed of sound is that using backscattering techniques there is not needed to know the thickness of the sample. Note using a single transducer the estimation of the thickness of the sample is critical to accurately estimate SOS and, therefore, scattering techniques might be preferred. This is a significant benefit to envisaged implant applications \textit{in-vivo} where the GBR can be monitored.

\section*{Acknowledgment}
This work has been developed within the framework of
the IVIO-UPV Chair, of the research project of the Universitat Polit\`ecnica de Val\`encia
``Desarrollo de tecnolog\'ias aplicadas al campo de la odontolog\'ia''
and is the result of the collaboration of researchers from the
Universitat Polit\`ecnica de Val\`encia and the Instituto Valenciano de Investigaciones Odontol\'ogicas (IVIO) in the field of dental
implantology with ultrasonic technology. This work has also been supported by UPV and FISABIO through the project OSEODENT (POLISABIO 2018).  The authors acknowledge financial support from European Union and Generalitat Valenciana through the European Regional Development Fund program (IDIFEDER/2018/022) and Ag\`encia Valenciana de la Innovaci\'o through Unitat Cient\'ifica d’Innovaci\'o Empresarial (INNCON00/18/9).
N.J. acknowledges financial support from Generalitat Valenciana through grant
APOSTD/2017/042.


\begin{thebibliography}{32}%
	\makeatletter
	\providecommand \@ifxundefined [1]{%
		\@ifx{#1\undefined}
	}%
	\providecommand \@ifnum [1]{%
		\ifnum #1\expandafter \@firstoftwo
		\else \expandafter \@secondoftwo
		\fi
	}%
	\providecommand \@ifx [1]{%
		\ifx #1\expandafter \@firstoftwo
		\else \expandafter \@secondoftwo
		\fi
	}%
	\providecommand \natexlab [1]{#1}%
	\providecommand \enquote  [1]{``#1''}%
	\providecommand \bibnamefont  [1]{#1}%
	\providecommand \bibfnamefont [1]{#1}%
	\providecommand \citenamefont [1]{#1}%
	\providecommand \href@noop [0]{\@secondoftwo}%
	\providecommand \href [0]{\begingroup \@sanitize@url \@href}%
	\providecommand \@href[1]{\@@startlink{#1}\@@href}%
	\providecommand \@@href[1]{\endgroup#1\@@endlink}%
	\providecommand \@sanitize@url [0]{\catcode `\\12\catcode `\$12\catcode
		`\&12\catcode `\#12\catcode `\^12\catcode `\_12\catcode `\%12\relax}%
	\providecommand \@@startlink[1]{}%
	\providecommand \@@endlink[0]{}%
	\providecommand \url  [0]{\begingroup\@sanitize@url \@url }%
	\providecommand \@url [1]{\endgroup\@href {#1}{\urlprefix }}%
	\providecommand \urlprefix  [0]{URL }%
	\providecommand \Eprint [0]{\href }%
	\providecommand \doibase [0]{http://dx.doi.org/}%
	\providecommand \selectlanguage [0]{\@gobble}%
	\providecommand \bibinfo  [0]{\@secondoftwo}%
	\providecommand \bibfield  [0]{\@secondoftwo}%
	\providecommand \translation [1]{[#1]}%
	\providecommand \BibitemOpen [0]{}%
	\providecommand \bibitemStop [0]{}%
	\providecommand \bibitemNoStop [0]{.\EOS\space}%
	\providecommand \EOS [0]{\spacefactor3000\relax}%
	\providecommand \BibitemShut  [1]{\csname bibitem#1\endcsname}%
	\let\auto@bib@innerbib\@empty
	\bibitem [{\citenamefont {Bodic}\ \emph {et~al.}(2005)\citenamefont {Bodic},
		\citenamefont {Hamel}, \citenamefont {Lerouxel}, \citenamefont {Basl{\'e}},\
		and\ \citenamefont {Chappard}}]{Bodic2005}%
	\BibitemOpen
	\bibfield  {author} {\bibinfo {author} {\bibfnamefont {F.}~\bibnamefont
			{Bodic}}, \bibinfo {author} {\bibfnamefont {L.}~\bibnamefont {Hamel}},
		\bibinfo {author} {\bibfnamefont {E.}~\bibnamefont {Lerouxel}}, \bibinfo
		{author} {\bibfnamefont {M.~F.}\ \bibnamefont {Basl{\'e}}}, \ and\ \bibinfo
		{author} {\bibfnamefont {D.}~\bibnamefont {Chappard}},\ }\bibfield  {title}
	{\enquote {\bibinfo {title} {Bone loss and teeth},}\ }\href@noop {}
	{\bibfield  {journal} {\bibinfo  {journal} {Joint Bone Spine}\ }\textbf
		{\bibinfo {volume} {72}},\ \bibinfo {pages} {215--221} (\bibinfo {year}
		{2005})}\BibitemShut {NoStop}%
	\bibitem [{\citenamefont {Tatum}(1986)}]{Tatum1986}%
	\BibitemOpen
	\bibfield  {author} {\bibinfo {author} {\bibfnamefont {J.~H.}\ \bibnamefont
			{Tatum}},\ }\bibfield  {title} {\enquote {\bibinfo {title} {Maxillary and
				sinus implant reconstructions.}}\ }\href@noop {} {\bibfield  {journal}
		{\bibinfo  {journal} {Dental Clinics of North America}\ }\textbf {\bibinfo
			{volume} {30}},\ \bibinfo {pages} {207--229} (\bibinfo {year}
		{1986})}\BibitemShut {NoStop}%
	\bibitem [{\citenamefont {Annibali}\ \emph {et~al.}(2008)\citenamefont
		{Annibali}, \citenamefont {Ripari}, \citenamefont {La~Monaca}, \citenamefont
		{Tonoli},\ and\ \citenamefont {Cristalli}}]{Annibali2008}%
	\BibitemOpen
	\bibfield  {author} {\bibinfo {author} {\bibfnamefont {S.}~\bibnamefont
			{Annibali}}, \bibinfo {author} {\bibfnamefont {M.}~\bibnamefont {Ripari}},
		\bibinfo {author} {\bibfnamefont {G.}~\bibnamefont {La~Monaca}}, \bibinfo
		{author} {\bibfnamefont {F.}~\bibnamefont {Tonoli}}, \ and\ \bibinfo {author}
		{\bibfnamefont {M.~P.}\ \bibnamefont {Cristalli}},\ }\bibfield  {title}
	{\enquote {\bibinfo {title} {Local complications in dental implant surgery:
				prevention and treatment},}\ }\href@noop {} {\bibfield  {journal} {\bibinfo
			{journal} {ORAL \& implantology}\ }\textbf {\bibinfo {volume} {1}},\ \bibinfo
		{pages} {21} (\bibinfo {year} {2008})}\BibitemShut {NoStop}%
	\bibitem [{\citenamefont {Acero}(2011)}]{JAcero2011}%
	\BibitemOpen
	\bibfield  {author} {\bibinfo {author} {\bibfnamefont {J.}~\bibnamefont
			{Acero}},\ }\bibfield  {title} {\enquote {\bibinfo {title} {Maxillary sinus
				grafting for implant insertion},}\ }in\ \href@noop {} {\emph {\bibinfo
			{booktitle} {Preprosthetic and Maxillofacial Surgery}}}\ (\bibinfo
	{publisher} {Elsevier},\ \bibinfo {year} {2011})\ pp.\ \bibinfo {pages}
	{54--75}\BibitemShut {NoStop}%
	\bibitem [{\citenamefont {Memon}\ \emph {et~al.}(2010)\citenamefont {Memon},
		\citenamefont {Godward}, \citenamefont {Williams}, \citenamefont {Siddique},\
		and\ \citenamefont {Al-Saleh}}]{memon2010dental}%
	\BibitemOpen
	\bibfield  {author} {\bibinfo {author} {\bibfnamefont {A.}~\bibnamefont
			{Memon}}, \bibinfo {author} {\bibfnamefont {S.}~\bibnamefont {Godward}},
		\bibinfo {author} {\bibfnamefont {D.}~\bibnamefont {Williams}}, \bibinfo
		{author} {\bibfnamefont {I.}~\bibnamefont {Siddique}}, \ and\ \bibinfo
		{author} {\bibfnamefont {K.}~\bibnamefont {Al-Saleh}},\ }\bibfield  {title}
	{\enquote {\bibinfo {title} {Dental x-rays and the risk of thyroid cancer: a
				case-control study},}\ }\href@noop {} {\bibfield  {journal} {\bibinfo
			{journal} {Acta Oncologica}\ }\textbf {\bibinfo {volume} {49}},\ \bibinfo
		{pages} {447--453} (\bibinfo {year} {2010})}\BibitemShut {NoStop}%
	\bibitem [{\citenamefont {Ludlow}, \citenamefont {Davies-Ludlow},\ and\
		\citenamefont {White}(2008)}]{Ludlow2008}%
	\BibitemOpen
	\bibfield  {author} {\bibinfo {author} {\bibfnamefont {J.~B.}\ \bibnamefont
			{Ludlow}}, \bibinfo {author} {\bibfnamefont {L.~E.}\ \bibnamefont
			{Davies-Ludlow}}, \ and\ \bibinfo {author} {\bibfnamefont {S.~C.}\
			\bibnamefont {White}},\ }\bibfield  {title} {\enquote {\bibinfo {title}
			{Patient risk related to common dental radiographic examinations: the impact
				of 2007 international commission on radiological protection recommendations
				regarding dose calculation},}\ }\href@noop {} {\bibfield  {journal} {\bibinfo
			{journal} {The Journal of the American Dental Association}\ }\textbf
		{\bibinfo {volume} {139}},\ \bibinfo {pages} {1237--1243} (\bibinfo {year}
		{2008})}\BibitemShut {NoStop}%
	\bibitem [{\citenamefont {Abendschein}\ and\ \citenamefont
		{Hyatt}(1970)}]{Abendschein1970}%
	\BibitemOpen
	\bibfield  {author} {\bibinfo {author} {\bibfnamefont {W.}~\bibnamefont
			{Abendschein}}\ and\ \bibinfo {author} {\bibfnamefont {G.}~\bibnamefont
			{Hyatt}},\ }\bibfield  {title} {\enquote {\bibinfo {title} {33 ultrasonics
				and selected physical properties of bone},}\ }\href@noop {} {\bibfield
		{journal} {\bibinfo  {journal} {Clinical Orthopaedics and Related Research}\
		}\textbf {\bibinfo {volume} {69}},\ \bibinfo {pages} {294--301} (\bibinfo
		{year} {1970})}\BibitemShut {NoStop}%
	\bibitem [{\citenamefont {Rossman}\ \emph {et~al.}(1989)\citenamefont
		{Rossman}, \citenamefont {Zagzebski}, \citenamefont {Mesina}, \citenamefont
		{Sorenson},\ and\ \citenamefont {Mazess}}]{Rossman1989}%
	\BibitemOpen
	\bibfield  {author} {\bibinfo {author} {\bibfnamefont {P.}~\bibnamefont
			{Rossman}}, \bibinfo {author} {\bibfnamefont {J.}~\bibnamefont {Zagzebski}},
		\bibinfo {author} {\bibfnamefont {C.}~\bibnamefont {Mesina}}, \bibinfo
		{author} {\bibfnamefont {J.}~\bibnamefont {Sorenson}}, \ and\ \bibinfo
		{author} {\bibfnamefont {R.}~\bibnamefont {Mazess}},\ }\bibfield  {title}
	{\enquote {\bibinfo {title} {Comparison of speed of sound and ultrasound
				attenuation in the os calcis to bone density of the radius, femur and lumbar
				spine},}\ }\href@noop {} {\bibfield  {journal} {\bibinfo  {journal} {Clinical
				Physics and Physiological Measurement}\ }\textbf {\bibinfo {volume} {10}},\
		\bibinfo {pages} {353} (\bibinfo {year} {1989})}\BibitemShut {NoStop}%
	\bibitem [{\citenamefont {Langton}, \citenamefont {Palmer},\ and\ \citenamefont
		{Porter}(1984)}]{Langton1984}%
	\BibitemOpen
	\bibfield  {author} {\bibinfo {author} {\bibfnamefont {C.}~\bibnamefont
			{Langton}}, \bibinfo {author} {\bibfnamefont {S.}~\bibnamefont {Palmer}}, \
		and\ \bibinfo {author} {\bibfnamefont {R.}~\bibnamefont {Porter}},\
	}\bibfield  {title} {\enquote {\bibinfo {title} {The measurement of broadband
				ultrasonic attenuation in cancellous bone},}\ }\href@noop {} {\bibfield
		{journal} {\bibinfo  {journal} {Engineering in Medicine}\ }\textbf {\bibinfo
			{volume} {13}},\ \bibinfo {pages} {89--91} (\bibinfo {year}
		{1984})}\BibitemShut {NoStop}%
	\bibitem [{\citenamefont {Langton}\ and\ \citenamefont
		{Njeh}(2008)}]{Langton2008}%
	\BibitemOpen
	\bibfield  {author} {\bibinfo {author} {\bibfnamefont {C.~M.}\ \bibnamefont
			{Langton}}\ and\ \bibinfo {author} {\bibfnamefont {C.~F.}\ \bibnamefont
			{Njeh}},\ }\bibfield  {title} {\enquote {\bibinfo {title} {The measurement of
				broadband ultrasonic attenuation in cancellous bone-a review of the science
				and technology},}\ }\href@noop {} {\bibfield  {journal} {\bibinfo  {journal}
			{IEEE Transactions on Ultrasonics, Ferroelectrics, and Frequency Control}\
		}\textbf {\bibinfo {volume} {55}},\ \bibinfo {pages} {1546--1554} (\bibinfo
		{year} {2008})}\BibitemShut {NoStop}%
	\bibitem [{\citenamefont {Laugier}\ and\ \citenamefont
		{Ha{\"\i}at}(2011)}]{Laugier2011}%
	\BibitemOpen
	\bibfield  {author} {\bibinfo {author} {\bibfnamefont {P.}~\bibnamefont
			{Laugier}}\ and\ \bibinfo {author} {\bibfnamefont {G.}~\bibnamefont
			{Ha{\"\i}at}},\ }\href@noop {} {\emph {\bibinfo {title} {Bone quantitative
				ultrasound}}},\ Vol.\ \bibinfo {volume} {576}\ (\bibinfo  {publisher}
	{Springer},\ \bibinfo {year} {2011})\BibitemShut {NoStop}%
	\bibitem [{\citenamefont {Wear}\ and\ \citenamefont {Garra}(1998)}]{Wear1998}%
	\BibitemOpen
	\bibfield  {author} {\bibinfo {author} {\bibfnamefont {K.~A.}\ \bibnamefont
			{Wear}}\ and\ \bibinfo {author} {\bibfnamefont {B.~S.}\ \bibnamefont
			{Garra}},\ }\bibfield  {title} {\enquote {\bibinfo {title} {Assessment of
				bone density using ultrasonic backscatter},}\ }\href@noop {} {\bibfield
		{journal} {\bibinfo  {journal} {Ultrasound in Medicine and Biology}\ }\textbf
		{\bibinfo {volume} {24}},\ \bibinfo {pages} {689--695} (\bibinfo {year}
		{1998})}\BibitemShut {NoStop}%
	\bibitem [{\citenamefont {Chaffa{\i}}\ \emph {et~al.}(2002)\citenamefont
		{Chaffa{\i}}, \citenamefont {Peyrin}, \citenamefont {Nuzzo}, \citenamefont
		{Porcher}, \citenamefont {Berger},\ and\ \citenamefont
		{Laugier}}]{Chaffai2002}%
	\BibitemOpen
	\bibfield  {author} {\bibinfo {author} {\bibfnamefont {S.}~\bibnamefont
			{Chaffa{\i}}}, \bibinfo {author} {\bibfnamefont {F.}~\bibnamefont {Peyrin}},
		\bibinfo {author} {\bibfnamefont {S.}~\bibnamefont {Nuzzo}}, \bibinfo
		{author} {\bibfnamefont {R.}~\bibnamefont {Porcher}}, \bibinfo {author}
		{\bibfnamefont {G.}~\bibnamefont {Berger}}, \ and\ \bibinfo {author}
		{\bibfnamefont {P.}~\bibnamefont {Laugier}},\ }\bibfield  {title} {\enquote
		{\bibinfo {title} {Ultrasonic characterization of human cancellous bone using
				transmission and backscatter measurements: relationships to density and
				microstructure},}\ }\href@noop {} {\bibfield  {journal} {\bibinfo  {journal}
			{Bone}\ }\textbf {\bibinfo {volume} {30}},\ \bibinfo {pages} {229--237}
		(\bibinfo {year} {2002})}\BibitemShut {NoStop}%
	\bibitem [{\citenamefont {Padilla}\ \emph {et~al.}(2008)\citenamefont
		{Padilla}, \citenamefont {Jenson}, \citenamefont {Bousson}, \citenamefont
		{Peyrin},\ and\ \citenamefont {Laugier}}]{padilla2008}%
	\BibitemOpen
	\bibfield  {author} {\bibinfo {author} {\bibfnamefont {F.}~\bibnamefont
			{Padilla}}, \bibinfo {author} {\bibfnamefont {F.}~\bibnamefont {Jenson}},
		\bibinfo {author} {\bibfnamefont {V.}~\bibnamefont {Bousson}}, \bibinfo
		{author} {\bibfnamefont {F.}~\bibnamefont {Peyrin}}, \ and\ \bibinfo {author}
		{\bibfnamefont {P.}~\bibnamefont {Laugier}},\ }\bibfield  {title} {\enquote
		{\bibinfo {title} {Relationships of trabecular bone structure with
				quantitative ultrasound parameters: In vitro study on human proximal femur
				using transmission and backscatter measurements},}\ }\href@noop {} {\bibfield
		{journal} {\bibinfo  {journal} {Bone}\ }\textbf {\bibinfo {volume} {42}},\
		\bibinfo {pages} {1193--1202} (\bibinfo {year} {2008})}\BibitemShut {NoStop}%
	\bibitem [{\citenamefont {Karjalainen}\ \emph {et~al.}(2009)\citenamefont
		{Karjalainen}, \citenamefont {T{\"o}yr{\"a}s}, \citenamefont {Riekkinen},
		\citenamefont {Hakulinen},\ and\ \citenamefont {Jurvelin}}]{karjalainen2009}%
	\BibitemOpen
	\bibfield  {author} {\bibinfo {author} {\bibfnamefont {J.~P.}\ \bibnamefont
			{Karjalainen}}, \bibinfo {author} {\bibfnamefont {J.}~\bibnamefont
			{T{\"o}yr{\"a}s}}, \bibinfo {author} {\bibfnamefont {O.}~\bibnamefont
			{Riekkinen}}, \bibinfo {author} {\bibfnamefont {M.}~\bibnamefont
			{Hakulinen}}, \ and\ \bibinfo {author} {\bibfnamefont {J.~S.}\ \bibnamefont
			{Jurvelin}},\ }\bibfield  {title} {\enquote {\bibinfo {title} {Ultrasound
				backscatter imaging provides frequency-dependent information on structure,
				composition and mechanical properties of human trabecular bone},}\
	}\href@noop {} {\bibfield  {journal} {\bibinfo  {journal} {Ultrasound in
				Medicine and Biology}\ }\textbf {\bibinfo {volume} {35}},\ \bibinfo {pages}
		{1376--1384} (\bibinfo {year} {2009})}\BibitemShut {NoStop}%
	\bibitem [{\citenamefont {Hoffmeister}\ \emph {et~al.}(2006)\citenamefont
		{Hoffmeister}, \citenamefont {Jones~III}, \citenamefont {Caldwell},\ and\
		\citenamefont {Kaste}}]{Hoffmeister2006}%
	\BibitemOpen
	\bibfield  {author} {\bibinfo {author} {\bibfnamefont {B.~K.}\ \bibnamefont
			{Hoffmeister}}, \bibinfo {author} {\bibfnamefont {C.}~\bibnamefont
			{Jones~III}}, \bibinfo {author} {\bibfnamefont {G.}~\bibnamefont {Caldwell}},
		\ and\ \bibinfo {author} {\bibfnamefont {S.}~\bibnamefont {Kaste}},\
	}\bibfield  {title} {\enquote {\bibinfo {title} {Ultrasonic characterization
				of cancellous bone using apparent integrated backscatter},}\ }\href@noop {}
	{\bibfield  {journal} {\bibinfo  {journal} {Physics in Medicine \& Biology}\
		}\textbf {\bibinfo {volume} {51}},\ \bibinfo {pages} {2715} (\bibinfo {year}
		{2006})}\BibitemShut {NoStop}%
	\bibitem [{\citenamefont {Hoffmeister}, \citenamefont {Holt},\ and\
		\citenamefont {Kaste}(2011)}]{Hoffmeister2011}%
	\BibitemOpen
	\bibfield  {author} {\bibinfo {author} {\bibfnamefont {B.~K.}\ \bibnamefont
			{Hoffmeister}}, \bibinfo {author} {\bibfnamefont {A.~P.}\ \bibnamefont
			{Holt}}, \ and\ \bibinfo {author} {\bibfnamefont {S.~C.}\ \bibnamefont
			{Kaste}},\ }\bibfield  {title} {\enquote {\bibinfo {title} {Effect of the
				cortex on ultrasonic backscatter measurements of cancellous bone},}\
	}\href@noop {} {\bibfield  {journal} {\bibinfo  {journal} {Physics in
				Medicine \& Biology}\ }\textbf {\bibinfo {volume} {56}},\ \bibinfo {pages}
		{6243} (\bibinfo {year} {2011})}\BibitemShut {NoStop}%
	\bibitem [{\citenamefont {Hoffmeister}\ \emph {et~al.}(2012)\citenamefont
		{Hoffmeister}, \citenamefont {Wilson}, \citenamefont {Gilbert},\ and\
		\citenamefont {Sellers}}]{Hoffmeister2012}%
	\BibitemOpen
	\bibfield  {author} {\bibinfo {author} {\bibfnamefont {B.~K.}\ \bibnamefont
			{Hoffmeister}}, \bibinfo {author} {\bibfnamefont {A.~R.}\ \bibnamefont
			{Wilson}}, \bibinfo {author} {\bibfnamefont {M.~J.}\ \bibnamefont {Gilbert}},
		\ and\ \bibinfo {author} {\bibfnamefont {M.~E.}\ \bibnamefont {Sellers}},\
	}\bibfield  {title} {\enquote {\bibinfo {title} {A backscatter difference
				technique for ultrasonic bone assessment},}\ }\href@noop {} {\bibfield
		{journal} {\bibinfo  {journal} {The Journal of the Acoustical Society of
				America}\ }\textbf {\bibinfo {volume} {132}},\ \bibinfo {pages} {4069--4076}
		(\bibinfo {year} {2012})}\BibitemShut {NoStop}%
	\bibitem [{\citenamefont {Naik}, \citenamefont {Malhotra},\ and\ \citenamefont
		{Popovics}(2003)}]{naik2003ultrasonic}%
	\BibitemOpen
	\bibfield  {author} {\bibinfo {author} {\bibfnamefont {T.~R.}\ \bibnamefont
			{Naik}}, \bibinfo {author} {\bibfnamefont {V.~M.}\ \bibnamefont {Malhotra}},
		\ and\ \bibinfo {author} {\bibfnamefont {J.~S.}\ \bibnamefont {Popovics}},\
	}\bibfield  {title} {\enquote {\bibinfo {title} {The ultrasonic pulse
				velocity method},}\ }in\ \href@noop {} {\emph {\bibinfo {booktitle} {Handbook
				on nondestructive testing of concrete}}}\ (\bibinfo  {publisher} {CRC
		Press},\ \bibinfo {year} {2003})\ pp.\ \bibinfo {pages}
	{182--200}\BibitemShut {NoStop}%
	\bibitem [{\citenamefont {Philippidis}\ and\ \citenamefont
		{Aggelis}(2005)}]{philippidis2005experimental}%
	\BibitemOpen
	\bibfield  {author} {\bibinfo {author} {\bibfnamefont {T.}~\bibnamefont
			{Philippidis}}\ and\ \bibinfo {author} {\bibfnamefont {D.}~\bibnamefont
			{Aggelis}},\ }\bibfield  {title} {\enquote {\bibinfo {title} {Experimental
				study of wave dispersion and attenuation in concrete},}\ }\href@noop {}
	{\bibfield  {journal} {\bibinfo  {journal} {Ultrasonics}\ }\textbf {\bibinfo
			{volume} {43}},\ \bibinfo {pages} {584--595} (\bibinfo {year}
		{2005})}\BibitemShut {NoStop}%
	\bibitem [{\citenamefont {Keating}, \citenamefont {Hannant},\ and\
		\citenamefont {Hibbert}(1989)}]{keating1989comparison}%
	\BibitemOpen
	\bibfield  {author} {\bibinfo {author} {\bibfnamefont {J.}~\bibnamefont
			{Keating}}, \bibinfo {author} {\bibfnamefont {D.}~\bibnamefont {Hannant}}, \
		and\ \bibinfo {author} {\bibfnamefont {A.}~\bibnamefont {Hibbert}},\
	}\bibfield  {title} {\enquote {\bibinfo {title} {Comparison of shear modulus
				and pulse velocity techniques to measure the build-up of structure in fresh
				cement pastes used in oil well cementing},}\ }\href@noop {} {\bibfield
		{journal} {\bibinfo  {journal} {Cement and Concrete Research}\ }\textbf
		{\bibinfo {volume} {19}},\ \bibinfo {pages} {554--566} (\bibinfo {year}
		{1989})}\BibitemShut {NoStop}%
	\bibitem [{\citenamefont {Sayers}\ and\ \citenamefont
		{Grenfell}(1993)}]{sayers1993ultrasonic}%
	\BibitemOpen
	\bibfield  {author} {\bibinfo {author} {\bibfnamefont {C.}~\bibnamefont
			{Sayers}}\ and\ \bibinfo {author} {\bibfnamefont {R.}~\bibnamefont
			{Grenfell}},\ }\bibfield  {title} {\enquote {\bibinfo {title} {Ultrasonic
				propagation through hydrating cements},}\ }\href@noop {} {\bibfield
		{journal} {\bibinfo  {journal} {Ultrasonics}\ }\textbf {\bibinfo {volume}
			{31}},\ \bibinfo {pages} {147--153} (\bibinfo {year} {1993})}\BibitemShut
	{NoStop}%
	\bibitem [{\citenamefont {Boumiz}, \citenamefont {Vernet},\ and\ \citenamefont
		{Tenoudji}(1996)}]{boumiz1996mechanical}%
	\BibitemOpen
	\bibfield  {author} {\bibinfo {author} {\bibfnamefont {A.}~\bibnamefont
			{Boumiz}}, \bibinfo {author} {\bibfnamefont {C.}~\bibnamefont {Vernet}}, \
		and\ \bibinfo {author} {\bibfnamefont {F.~C.}\ \bibnamefont {Tenoudji}},\
	}\bibfield  {title} {\enquote {\bibinfo {title} {Mechanical properties of
				cement pastes and mortars at early ages: Evolution with time and degree of
				hydration},}\ }\href@noop {} {\bibfield  {journal} {\bibinfo  {journal}
			{Advanced cement based materials}\ }\textbf {\bibinfo {volume} {3}},\
		\bibinfo {pages} {94--106} (\bibinfo {year} {1996})}\BibitemShut {NoStop}%
	\bibitem [{\citenamefont {Carlson}\ \emph {et~al.}(2003)\citenamefont
		{Carlson}, \citenamefont {Nilsson}, \citenamefont {Fern{\'a}ndez},\ and\
		\citenamefont {Planell}}]{Carlson2003}%
	\BibitemOpen
	\bibfield  {author} {\bibinfo {author} {\bibfnamefont {J.}~\bibnamefont
			{Carlson}}, \bibinfo {author} {\bibfnamefont {M.}~\bibnamefont {Nilsson}},
		\bibinfo {author} {\bibfnamefont {E.}~\bibnamefont {Fern{\'a}ndez}}, \ and\
		\bibinfo {author} {\bibfnamefont {J.}~\bibnamefont {Planell}},\ }\bibfield
	{title} {\enquote {\bibinfo {title} {An ultrasonic pulse-echo technique for
				monitoring the setting of caso4-based bone cement},}\ }\href@noop {}
	{\bibfield  {journal} {\bibinfo  {journal} {Biomaterials}\ }\textbf {\bibinfo
			{volume} {24}},\ \bibinfo {pages} {71--77} (\bibinfo {year}
		{2003})}\BibitemShut {NoStop}%
	\bibitem [{\citenamefont {Vlad}\ \emph {et~al.}(2012)\citenamefont {Vlad},
		\citenamefont {Gonz{\'a}lez}, \citenamefont {G{\'o}mez}, \citenamefont
		{L{\'o}pez}, \citenamefont {Carlson},\ and\ \citenamefont
		{Fern{\'a}ndez}}]{vlad2012ultrasound}%
	\BibitemOpen
	\bibfield  {author} {\bibinfo {author} {\bibfnamefont {M.}~\bibnamefont
			{Vlad}}, \bibinfo {author} {\bibfnamefont {L.}~\bibnamefont {Gonz{\'a}lez}},
		\bibinfo {author} {\bibfnamefont {S.}~\bibnamefont {G{\'o}mez}}, \bibinfo
		{author} {\bibfnamefont {J.}~\bibnamefont {L{\'o}pez}}, \bibinfo {author}
		{\bibfnamefont {J.~E.}\ \bibnamefont {Carlson}}, \ and\ \bibinfo {author}
		{\bibfnamefont {E.}~\bibnamefont {Fern{\'a}ndez}},\ }\bibfield  {title}
	{\enquote {\bibinfo {title} {Ultrasound monitoring of the setting of
				calcium-based bone cements},}\ }\href@noop {} {\bibfield  {journal} {\bibinfo
			{journal} {Journal of Materials Science: Materials in Medicine}\ }\textbf
		{\bibinfo {volume} {23}},\ \bibinfo {pages} {1563--1568} (\bibinfo {year}
		{2012})}\BibitemShut {NoStop}%
	\bibitem [{\citenamefont {Finkemeier}(2002)}]{finkemeier2002bone}%
	\BibitemOpen
	\bibfield  {author} {\bibinfo {author} {\bibfnamefont {C.~G.}\ \bibnamefont
			{Finkemeier}},\ }\bibfield  {title} {\enquote {\bibinfo {title}
			{Bone-grafting and bone-graft substitutes},}\ }\href@noop {} {\bibfield
		{journal} {\bibinfo  {journal} {The Journal of Bone and Joint Surgery}\
		}\textbf {\bibinfo {volume} {84}},\ \bibinfo {pages} {454--464} (\bibinfo
		{year} {2002})}\BibitemShut {NoStop}%
	\bibitem [{\citenamefont {Wang}\ and\ \citenamefont
		{Yeung}(2017)}]{wang2017bone}%
	\BibitemOpen
	\bibfield  {author} {\bibinfo {author} {\bibfnamefont {W.}~\bibnamefont
			{Wang}}\ and\ \bibinfo {author} {\bibfnamefont {K.~W.}\ \bibnamefont
			{Yeung}},\ }\bibfield  {title} {\enquote {\bibinfo {title} {Bone grafts and
				biomaterials substitutes for bone defect repair: A review},}\ }\href@noop {}
	{\bibfield  {journal} {\bibinfo  {journal} {Bioactive Materials}\ }\textbf
		{\bibinfo {volume} {2}},\ \bibinfo {pages} {224--247} (\bibinfo {year}
		{2017})}\BibitemShut {NoStop}%
	\bibitem [{\citenamefont {De~Leonardis}\ and\ \citenamefont
		{Pecora}(2000)}]{DeLeonardis2000}%
	\BibitemOpen
	\bibfield  {author} {\bibinfo {author} {\bibfnamefont {D.}~\bibnamefont
			{De~Leonardis}}\ and\ \bibinfo {author} {\bibfnamefont {G.~E.}\ \bibnamefont
			{Pecora}},\ }\bibfield  {title} {\enquote {\bibinfo {title} {Prospective
				study on the augmentation of the maxillary sinus with calcium sulfate:
				histological results},}\ }\href@noop {} {\bibfield  {journal} {\bibinfo
			{journal} {Journal of Periodontology}\ }\textbf {\bibinfo {volume} {71}},\
		\bibinfo {pages} {940--947} (\bibinfo {year} {2000})}\BibitemShut {NoStop}%
	\bibitem [{\citenamefont {Thomas}\ and\ \citenamefont
		{Puleo}(2009)}]{Thomas2009}%
	\BibitemOpen
	\bibfield  {author} {\bibinfo {author} {\bibfnamefont {M.~V.}\ \bibnamefont
			{Thomas}}\ and\ \bibinfo {author} {\bibfnamefont {D.~A.}\ \bibnamefont
			{Puleo}},\ }\bibfield  {title} {\enquote {\bibinfo {title} {Calcium sulfate:
				Properties and clinical applications},}\ }\href@noop {} {\bibfield  {journal}
		{\bibinfo  {journal} {Journal of Biomedical Materials Research Part B:
				Applied Biomaterials: An Official Journal of The Society for Biomaterials,
				The Japanese Society for Biomaterials, and The Australian Society for
				Biomaterials and the Korean Society for Biomaterials}\ }\textbf {\bibinfo
			{volume} {88}},\ \bibinfo {pages} {597--610} (\bibinfo {year}
		{2009})}\BibitemShut {NoStop}%
	\bibitem [{\citenamefont {Hoffmeister}\ \emph {et~al.}(2015)\citenamefont
		{Hoffmeister}, \citenamefont {Mcpherson}, \citenamefont {Smathers},
		\citenamefont {Spinolo},\ and\ \citenamefont {Sellers}}]{Hoffmeister2015}%
	\BibitemOpen
	\bibfield  {author} {\bibinfo {author} {\bibfnamefont {B.~K.}\ \bibnamefont
			{Hoffmeister}}, \bibinfo {author} {\bibfnamefont {J.~A.}\ \bibnamefont
			{Mcpherson}}, \bibinfo {author} {\bibfnamefont {M.~R.}\ \bibnamefont
			{Smathers}}, \bibinfo {author} {\bibfnamefont {P.~L.}\ \bibnamefont
			{Spinolo}}, \ and\ \bibinfo {author} {\bibfnamefont {M.~E.}\ \bibnamefont
			{Sellers}},\ }\bibfield  {title} {\enquote {\bibinfo {title} {Ultrasonic
				backscatter from cancellous bone: The apparent backscatter transfer
				function},}\ }\href@noop {} {\bibfield  {journal} {\bibinfo  {journal} {IEEE
				Transactions on Ultrasonics, Ferroelectrics, and Frequency Control}\ }\textbf
		{\bibinfo {volume} {62}},\ \bibinfo {pages} {2115--2125} (\bibinfo {year}
		{2015})}\BibitemShut {NoStop}%
	\bibitem [{\citenamefont {Liu}\ \emph {et~al.}(2003)\citenamefont {Liu},
		\citenamefont {Gai}, \citenamefont {Pan},\ and\ \citenamefont
		{Liu}}]{liu2003exothermal}%
	\BibitemOpen
	\bibfield  {author} {\bibinfo {author} {\bibfnamefont {C.}~\bibnamefont
			{Liu}}, \bibinfo {author} {\bibfnamefont {W.}~\bibnamefont {Gai}}, \bibinfo
		{author} {\bibfnamefont {S.}~\bibnamefont {Pan}}, \ and\ \bibinfo {author}
		{\bibfnamefont {Z.}~\bibnamefont {Liu}},\ }\bibfield  {title} {\enquote
		{\bibinfo {title} {The exothermal behavior in the hydration process of
				calcium phosphate cement},}\ }\href@noop {} {\bibfield  {journal} {\bibinfo
			{journal} {Biomaterials}\ }\textbf {\bibinfo {volume} {24}},\ \bibinfo
		{pages} {2995--3003} (\bibinfo {year} {2003})}\BibitemShut {NoStop}%
	\bibitem [{\citenamefont {Ester}\ \emph {et~al.}(1996)\citenamefont {Ester},
		\citenamefont {Kriegel}, \citenamefont {Sander}, \citenamefont {Xu} \emph
		{et~al.}}]{ester1996density}%
	\BibitemOpen
	\bibfield  {author} {\bibinfo {author} {\bibfnamefont {M.}~\bibnamefont
			{Ester}}, \bibinfo {author} {\bibfnamefont {H.-P.}\ \bibnamefont {Kriegel}},
		\bibinfo {author} {\bibfnamefont {J.}~\bibnamefont {Sander}}, \bibinfo
		{author} {\bibfnamefont {X.}~\bibnamefont {Xu}},  \emph {et~al.},\ }\bibfield
	{title} {\enquote {\bibinfo {title} {A density-based algorithm for
				discovering clusters in large spatial databases with noise.}}\ }in\
	\href@noop {} {\emph {\bibinfo {booktitle} {Kdd}}},\ Vol.~\bibinfo {volume}
	{96}\ (\bibinfo {year} {1996})\ pp.\ \bibinfo {pages} {226--231}\BibitemShut
	{NoStop}%
\end{thebibliography}

%

%




\end{document}